\documentclass[aps,pre,reprint,groupedaddress,showpacs]{revtex4-1}
\usepackage{amsmath,amssymb,amsxtra,amsfonts,graphicx,wrapfig}
\usepackage{xcolor}
\usepackage{subfigure}
\usepackage{enumerate}
\usepackage[font=scriptsize,labelfont=bf]{caption}
\usepackage{color}
\begin{document}
\title{Steady-state dynamics of exclusion process with local reversible association of particles}
\author{A. Jindal$^1$}
\author{A.B. Kolomeisky$^2$}
\author{A.K. Gupta$^1$}
\email[]{akgupta@iitrpr.ac.in}
\affiliation{$^1$ Department of Mathematics, Indian Institute of Technology Ropar,
Rupnagar-140001, Punjab, India\\
$^2$ Department of Chemistry and Chemical and Biomolecular Engineering,
Rice University, Houston, TX 77005, United States of America}

\begin{abstract}

Many biological processes are supported by special molecules, called motor proteins or molecular motors, that transport cellular cargoes along linear protein filaments and can reversibly associate to their tracks. Stimulated by these observations, we developed a theoretical model for collective dynamics of biological molecular motors that accounts for local association/dissociation events. In our approach, the particles interacting only via exclusion move along a lattice in the preferred direction, while the reversible associations are allowed at the specific site far away from the boundaries. Considering the association/dissociation site as a local defect, the inhomogeneous system is approximated as two coupled homogeneous sub-lattices. This allows us to obtain a full description of stationary dynamics in the system. It is found that the number and nature of steady-state phases strongly depend on the values of association and dissociation transition rates. Microscopic arguments to explain these observations as well as biological implications are also discussed. Theoretical predictions agree well with extensive Monte Carlo computer simulations.

\end{abstract}

\maketitle

\section{Introduction}

Multiple cellular processes, such as cell division, cell locomotion, cell motility, and cellular cargo transport, are driven by several classes of biological molecules that are known as motor proteins or molecular motors \cite{howard2001mechanics,bray2000cell,kolomeisky2007molecular,chowdhury2013stochastic,kolomeisky2013motor,kolomeisky2015motor}. These are special enzymatic proteins that catalyze the hydrolysis of energy-rich adenosine triphosphate (ATP) or biopolymerization of nucleic acids and proteins \cite{kolomeisky2015motor}. The released chemical energy is then converted to mechanical energy that supports the movement of cargo-carrying motors on active biological filaments. Significant advances in understanding the mechanisms and single-molecule properties of various molecular motors have been achieved due to a large volume of experimental and theoretical investigations \cite{chowdhury2013stochastic,kolomeisky2013motor}. However in live cells, motor proteins typically operate in groups that interact with each other \cite{vilfan2001dynamics,roos2008dynamic,teimouri2015theoretical,midha2018effect,celis2015correlations}. But our knowledge of the microscopic mechanisms of collective dynamics of biological molecular motors are still very limited \cite{neri2013exclusion,kolomeisky2013motor}.

The most popular approach to investigate the multi-particle dynamics in low-dimensional non-equilibrium systems remain the exploration of so-called exclusion processes \cite{katz1983phase,chowdhury2005physics,blythe2007nonequilibrium}. To study the dynamics of directed molecular motors, a special class of driven diffusive models known as Totally Asymmetric Simple Exclusion Processes (TASEP) has been utilized to explore the microscopic system properties \cite{kolomeisky1998phase,macdonald1968kinetics,chou2011non,parmeggiani2003phase}. It was originally introduced in 1968 to study the kinetics of biopolymerization \cite{macdonald1968kinetics}, and since then, it has been further generalized to study the motion of molecular motors  and other biological transport processes \cite{chou2011non}. Various aspects of molecular motors have been extensively analyzed utilizing different versions of TASEP models. Several interesting phenomena, such as phase separation, phase segregation, and boundary induced phase transitions have been discovered in these studies \cite{krug1991boundary,derrida1993exact,schutz1993phase,kolomeisky1998phase}. In most cases, the stationary behavior of complex multi-particle non-equilibrium processes have been thoroughly analyzed using exact solutions or  mean-field approximations that neglect the correlations  between particles \cite{chou2011non,kolomeisky1998phase}.

One of the important features of biological molecular motors is their ability to reversibly associate from the linear tracks. This effect, when the probability of association/dissociation is the same for all sites on the filaments, has been rigorously explored in so-called TASEP models with Langmuir Kinetics \cite{parmeggiani2004totally}. It was also observed  in real systems that the reversible association of the motor proteins is not always  a homogeneous process, i.e., there are specific sites from which the motor proteins can preferentially dissociate or associate. Such situations have been theoretically investigated, but to a less degree \cite{xiao2012single,xiao2019effect,mirin2003effect}. Importantly, in these studies the associations and dissociations were not considered together, which limited the understanding of underlying microscopic processes in the motor proteins transport.

Stimulated by these observations, we develop a theoretical model for the collective transport of molecular motors that takes into account the possibility of reversible association/dissociation at a specific site. Our goal is to understand how the localized association/dissociation affects the collective dynamics. By noticing that the site where association/dissociation events are taking place divides the originally inhomogeneous system into two coupled homogeneous TASEPs, the steady-state particles dynamics is calculated. We explicitly determined how non-equilibrium phase diagrams and dynamic properties vary with changing the association and dissociation transition rates. Our theoretical results are supported by Monte Carlo computer simulations.

\begin{figure*}
\centering
\includegraphics[width=0.9\textwidth]{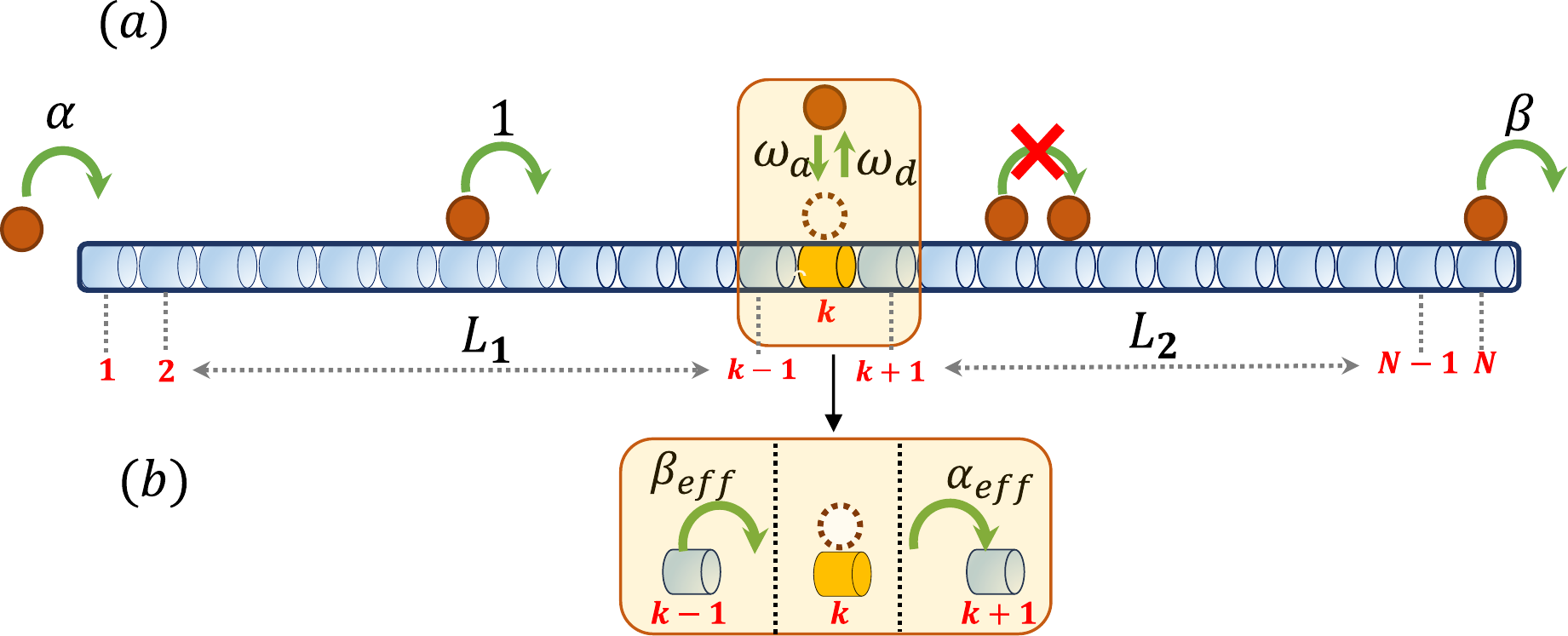}
\caption{\label{Model} (a) A schematic view of the TASEP model with localized associations/dissociations. Identical particles can enter the vacant first site with the rate $\alpha$ and escape the lattice  from the last site with the rate $\beta$. In the bulk, particles jump to the neighboring site with the unit rate. In addition, at the site $k$ particles are allowed to dissociate with the rate $\omega_d$, or the particle from the outside can associate to the empty site $k$ with the rate $\omega_a$. The lattice is divided into two homogeneous segments $L_1$ and $L_2$ coupled at special site $k$. (b) The particles can leave the left segment with effective exit rate $\beta_{eff}$ and enter into the right segment with effective entry rate $\alpha_{eff}$.}
\end{figure*}

\section{Model}

Motivated by the reversible association of molecular motors on biological filaments, we propose a one-dimensional TASEP model to analyse the properties of inhomogeneous non-equilibrium systems. In our model, the transport of motors is considered as a movement of particles along a lattice segment with $N$ discrete sites as presented in Fig. \ref{Model}(a). Sites $i=1$ and $i=N$ represent the left and right boundary, respectively, while the sites $1<i<N$ constitute the bulk of the lattice. At each site, particles obey a hard core exclusion principle that restricts the site to be occupied by no more than one particle. A particle is injected into the lattice through the left boundary with a rate $\alpha$, and it leaves from the right boundary with a rate $\beta$ as shown in Fig. \ref{Model}(a). The particles, which mimic the biological molecular motors, move in one preferred direction (to the right) in accordance to random sequential update rules. In the bulk, a particle at site $i$ is allowed to jump to the immediate site $i+1$ with a unit rate, provided that the site empty.

To take into account the localized reversible association/dissociation dynamics of molecular motors, we assume that this process can occur only at the special site $k=N/2$: see Fig. \ref{Model}(a). Since the filaments are typically very long, we consider a thermodynamic limit ($N \rightarrow \infty$), and in this case the exact location of the association/dissociation site does not affect the dynamics in the system as long as the special site is far away from the boundaries. A particle at the site $k$ can dissociate from the lattice with a rate $\omega_d$ and associate to the special site with a rate $\omega_a$, if this site is currently not occupied (see Fig. \ref{Model}(a)).

\begin{table*}
\caption{\label{simple}Summary of results for a simple homogeneous TASEP model \cite{derrida1992exact}.}
\begin{center}
\resizebox{0.5\textwidth}{!}{
\begin{tabular}{||c|c||c|c|c|c||}
\hline \hline
&Phase Region&$\rho_1$&$\rho_{bulk}$&$\rho_N$&Current($J$)\\
\hline \hline
LD&$\alpha<\min\{\beta,0.5\}$&$\alpha$&$\alpha$&$\frac{\alpha(1-\alpha)}{\beta}$&$\alpha(1-\alpha)$\\
\hline
HD&$\beta<\min\{\alpha,0.5\}$&$1-\frac{\beta(1-\beta)}{\alpha}$&$1-\beta$&$1-\beta$&$\beta(1-\beta)$\\
\hline
MC&$0.5<\min\{\alpha,\beta\}$&$1-\frac{1}{4\alpha}$&$0.5$&$\frac{1}{4\beta}$&$0.25$\\
\hline \hline
\end{tabular}
}
\end{center}
\end{table*}

\section{Theoretical Analysis}

The majority of investigated TASEP models analyze the homogeneous processes when the dynamics at all bulk sites is identical. This allowed researchers to obtain explicit description of dynamic properties of these systems \cite{kolomeisky1998phase,derrida1993exact,schutz1993phase,derrida1992exact}. Our model, however, deals with the inhomogeneous system due to the presence of the special site for reversible association/dissociation events. This significantly complicates the analysis. At the same time, we notice that the special site divides the originally inhomogeneous system into two coupled homogeneous sub-lattices: left segment $L_1$  ($i=1,2,\hdots,k-1$) and right segment $L_2$ ($i=k+1,k+2,\hdots,N$). This suggests that our model can be analyzed by considering it as two homogeneous TASEP lattices combined together by the special site as illustrated in Fig. \ref{Model}(a). Since the dynamics of exclusion processes on homogeneous lattices is fully quantized, this will help us to describe the inhomogeneous system. This is the main idea of our theoretical approach.

 One can define an effective exit rate of particles from the left segment $L_1$ as $\beta_{eff}$, and an effective entry rate of particles into the right segment $L_2$ as $\alpha_{eff}$ (see Fig.\ref{Model}(b)). Then, by utilizing stationary current arguments at the sites $k-1$, $k$ and $k+1$, we couple both the segments and explicitly calculate the effective transition rates. Furthermore, employing the results of homogeneous TASEP on two segments separately, our aim is to compute the effective rates and densities at the sites $k-1$, $k$ and $k+1$. We denote the average density of particles in the bulk of left and right segment as $\rho_{bulk,L_1}$ and $\rho_{bulk,L_2}$, respectively, and at the special sites as $\rho_i$ ($i=1,k-1,k,k+1,N$). The corresponding bulk current in $L_1$ and $L_2$ is represented by $J_{bulk,L_1}$ and $J_{bulk,L_2}$ respectively. Whereas, the current leaving the left segment is $J_{exit,L_1}$ and the current entering into the right segment is $J_{entry,L_2}$. In addition, the current associated with the association and dissociation of particles at the site $k$ is represented by $J_{a}$ and $J_{d}$, respectively.

%
%\begin{figure*}[!h]
%\centering
%\includegraphics[width=0.9\textwidth]{fig2}
%\caption{\label{break} Mapping of the original model into two coupled homogeneous segments $L_1$ and $L_2$. The particles can leave the left segment with effective exit rate $\beta_{eff}$ and enter into the right segment with effective entry rate $\alpha_{eff}$.}
%\end{figure*}

\subsection{Sub-Lattice Mean-Field Approximation}

The stationary properties of the open TASEP model with excluded volume interactions among particles have been explicitly obtained using various exact and approximate methods \cite{kolomeisky1998phase,derrida1993exact,schutz1993phase,derrida1992exact}. Importantly, the results of exact calculations mean-field approximations agree for the description of phase diagrams, particle current, and most particle densities. This allows us to use mean-field arguments in our derivations. It has been found that, depending on the values of entrance, exit and bulk hopping rates, there exist three distinct steady-state phases: entry dominated low-density (LD), exit dominated high-density (HD) and bulk dominated maximal-current (MC) \cite{derrida1992exact}. For convenience, the conditions of the existence, the particle densities, and the currents for the different phases obtained in simple open TASEP model i.e. $\omega_a=0$ and $\omega_d=0$ are summarized in Table \ref{simple}.

In the stationary state, the condition of the current conservation at the special site $k$ couples the fluxes in two segments as
\begin{equation}\label{coupling}
J_{a}+J_{pass,L_1}=J_{d}+J_{pass,L_2}
\end{equation}
where $J_{pass,L_1}$ denotes the  passing current from the site $k-1$ to $k$, and $J_{pass,L_2}$ denotes the passing current from the site $k$ to $k+1$. The expressions for these currents are given by,
\begin{eqnarray}\label{current1}
J_{a}=\omega_a(1-\rho_k),&&\qquad J_{d}=\omega_d\rho_k,\\\label{current2}
J_{pass,L_1}=\rho_{k-1}(1-\rho_k),&&\qquad J_{pass,L_2}=\rho_k(1-\rho_{k+1}).
\end{eqnarray}
Furthermore, the exit current from $L_1$ and the entry current into $L_2$ are given by,
\begin{equation}
J_{exit,L_1}=\beta_{eff}\rho_{k-1},\qquad J_{entry,L_2}=\alpha_{eff}(1-\rho_{k+1}).
\end{equation}
Also, from the stationarity of the current ($J_{pass,L_1}=J_{exit,L_1}$ and $J_{pass,L_2}=J_{entry,L_2}$) one could easily obtain,
\begin{equation}\label{betaeff}
\beta_{eff}=1-\rho_k, \quad \alpha_{eff}=\rho_k,
\end{equation}
which leads to
\begin{equation}\label{relation}
\alpha_{eff}=1-\beta_{eff}.
\end{equation}
In addition, we can rewrite Eqn. \eqref{coupling} as,
\begin{equation}\label{coupled}
\beta_{eff}(\omega_a+\rho_{k-1})=\alpha_{eff}(\omega_d+1-\rho_{k+1}).
\end{equation}
One could also notice that the current is constant throughout the homogeneous left and right segments separately, producing the following relations,
\begin{eqnarray}
J_{bulk,L_1}=\beta_{eff}\rho_{k-1},\qquad
J_{bulk,L_2}=\alpha_{eff}(1-\rho_{k+1}).
\end{eqnarray}

Now we can estimate the number of the stationary phases that might exist in our inhomogeneous model. Since each homogeneous segment can exhibit one of three possible phases (LD, HD or MC), there are $3^2=9$ possible phases in the system. However, not all phases might be realized due to the stationary condition on the particle current. To simplify our discussions, we label the possible phases as $A:B$, where $A$ and $B$ correspond to the phases in the left and right segments, respectively. Utilizing the results for open homogeneous TASEP model described in the Table \ref{simple}, the conditions of existence of different phases will be discussed in the next section.

\subsection{Stationary Phases and Explicit Phase Boundaries}\label{states}

Since each stationary state is explicitly characterized by specific parameters (see Table 1), we can determine the range of parameters for each possible steady-state dynamic regime.\\
{\it LD:LD Phase}. In this phase, both the segments $L_1$ and $L_2$ exhibit the low density phase that corresponds to
\begin{equation}\label{ldldcon}
\alpha\leq\min\{\beta_{eff},0.5\},\qquad \alpha_{eff}\leq\min\{\beta,0.5\}.
\end{equation}
The particle densities in the bulk of two segments are given by,
\begin{equation}
\rho_{buk,L_1}=\alpha,\qquad \rho_{bulk,L_2}=\alpha_{eff},
\end{equation}
while the particle density at the sites $k-1$ and $k+1$ are,
\begin{equation}\label{ldldden}
\rho_{k-1}=\frac{\alpha(1-\alpha)}{\beta_{eff}},\qquad \rho_{k+1}=\alpha_{eff}.
\end{equation}
To determine the explicit phase boundary, we need to compute the effective rates $\alpha_{eff}$ and $\beta_{eff}$. By substituting the expressions of $\rho_{k-1}$, $\rho_{k+1}$ from Eqn. \eqref{ldldden} into Eqn. \eqref{coupled} and using Eqn. \eqref{relation}, the effective rates are the following,
    \begin{eqnarray*}
    \alpha_{eff}&=&\frac{1}{2} \left(1+\omega_d+\omega_a-\sqrt{4 (\alpha^2-\alpha-\omega_a)+(\omega_d+\omega_a+1)^2}\right),\\
    \beta_{eff}&=&\frac{1}{2} \left(1-\omega_d-\omega_a+\sqrt{4 (\alpha^2-\alpha-\omega_a)+(\omega_d+\omega_a+1)^2}\right).
    \end{eqnarray*}
These explicit expressions together with Eq. (9) determine the range for the existence of the LD:LD phase. Also, one can easily conclude that this phase exists only when association/dissociation rates satisfy,
    \begin{equation}
    \omega_a-\omega_d\leq0.5.
    \end{equation}
Physically, this corresponds to the situation when the association is weaker or only slightly stronger than the dissociation process so that there is no substantial increase in the amount of particles entering into the system at the special site $k$.\\

{\it HD:HD Phase}. In this phase, both left and right segments are found in the high density phase with boundary parameters satisfying,
\begin{equation}\label{hdhdcon}
\beta_{eff}\leq\min\{\alpha,0.5\},\qquad \beta\leq\min\{\alpha_{eff},0.5\}.
\end{equation}
The bulk densities of the particles in each segment are
\begin{equation}
\rho_{buk,L_1}=1-\beta_{eff},\qquad \rho_{bulk,L_2}=1-\beta,
\end{equation}
while at the sites $k-1$ and $k+1$ we have
\begin{equation}\label{hdhdden}
\rho_{k-1}=1-\beta_{eff},\qquad \rho_{k+1}=1-\frac{\beta(1-\beta)}{\alpha_{eff}}.
\end{equation}
The effective rates $\alpha_{eff}$ and $\beta_{eff}$ can be computed by substituting the corresponding $\rho_{k-1}$ and $\rho_{k+1}$ from Eq. \eqref{hdhdden} into Eqn. \eqref{coupled} and utilizing Eq. \eqref{relation}, that yields,
       \begin{eqnarray*}
    \alpha_{eff}&=&\frac{1}{2} \left(1-\omega_d-\omega_a+\sqrt{(\omega_d+\omega_a-1)^2-4(\beta-\beta^2-\omega_a)}\right),\\
    \beta_{eff}&=&\frac{1}{2} \left(1+\omega_d+\omega_a-\sqrt{(\omega_d+\omega_a-1)^2-4(\beta-\beta^2-\omega_a)}\right).
    \end{eqnarray*}
These expressions together with Eq. \eqref{hdhdcon}  specify the
conditions for the existence of the HD:HD phase. Moreover, it can be shown that in this phase   the association and dissociation rates must satisfy
  \begin{equation}
    \omega_d-\omega_a\leq0.5.
    \end{equation}
Physically, this means that the phase exists when the dissociation is weaker or only slightly stronger than the association process so that the amount of particles leaving the system is relatively small.\\

{\it MC:MC Phase}. For this phase, the two segments $L_1$ and $L_2$ are assumed to be in maximal current phase specified by
\begin{equation}\label{mcmccon}
0.5\leq\min\{\alpha,\beta_{eff}\},\qquad 0.5\leq\min\{\alpha_{eff},\beta\}
\end{equation}
The bulk densities in the segments $L_1$ and $L_2$ are given by
\begin{equation}
\rho_{buk,L_1}=0.5,\qquad \rho_{bulk,L_2}=0.5,
\end{equation}
In addition, the densities at boundaries of two segments are equal to
\begin{equation}\label{mcmcden}
\rho_{k-1}=\frac{1}{4\beta_{eff}},\qquad \rho_{k+1}=1-\frac{1}{4\alpha_{eff}}.
\end{equation}
Then, using Eqs. \eqref{relation} and \eqref{coupled} we obtain the effective entrance and exit rates,
     \begin{equation}
    \alpha_{eff}=\frac{\omega_a}{\omega_a+\omega_d}, \quad \beta_{eff}=\frac{\omega_d}{\omega_a+\omega_d}.
    \end{equation}
For the effective rates $\alpha_{eff}$, $\beta_{eff}$ to satisfy Eq. \eqref{mcmccon}, we obtain
     \begin{equation}
     \omega_a=\omega_d, \quad \alpha_{eff}=\beta_{eff}=1/2.
     \end{equation}
The physical meaning of this result is that this phase exists only at the conditions when the association is always compensated by the dissociation ($J_{a}=J_{d}$) so that the overall system becomes fully homogeneous at all sites.\\

\begin{table*}
\caption{\label{limiting} Conditions for the existence of different stationary phase regimes in terms of association/dissociation rates where ``$\times$" denotes the phase that does not exist.}
\begin{center}
\resizebox{0.6\textwidth}{!}{
\begin{tabular}{c|c|c|c}
\hline \hline
Phase&$\omega_a=0,~\omega_d>0$&$\omega_d=0,~\omega_a>0$&$\omega_d>0,~\omega_a>0$\\
\hline \hline
LD:LD&$\omega_d>0$&$\omega_a\leq0.5$&$\omega_a-\omega_d\leq0.5$\\
HD:HD& $\omega_d\leq0.5$ &$\omega_a>0$&$\omega_d-\omega_a\leq0.5$\\
MC:MC & $\times$ &$\times$&$\omega_a=\omega_d$\\
LD:HD&$\omega_d>0$&$\omega_a>0$&$\omega_a,~\omega_d>0$\\
LD:MC &$\times$ &$\omega_a>0$&$\omega_a>\omega_d$\\
HD:MC &$\times$ & $\omega_a>0$&$\omega_a>\omega_d$\\
MC:HD &$\omega_d>0$ &$\times$&$\omega_a\leq\omega_d$\\
MC:LD &$\omega_d>0$ & $\times$&$\omega_a\leq\omega_d$\\
HD:LD & $\times$ &$\times$ &$\times$\\
\hline \hline
\end{tabular}
}
\end{center}
\end{table*}

{\it LD:MC Phase}. Here the left segment is in the LD phase, while the right segment is in the MC phase, and it is governed by the following conditions,
 \begin{equation}\label{ldmccon}
\alpha\leq\min\{\beta_{eff},0.5\},\qquad 0.5\leq\min\{\alpha_{eff},\beta\}
\end{equation}
The bulk densities  are given by,
\begin{equation}
\rho_{buk,L_1}=\alpha,\qquad \rho_{bulk,L_2}=0.5,
\end{equation}
while at the sites $k-1$ and $k+1$ we have
\begin{equation}\label{ldmcdemn}
\rho_{k-1}=\frac{\alpha(1-\alpha)}{\beta_{eff}},\qquad \rho_{k+1}=1-\frac{1}{4\alpha_{eff}}.
\end{equation}
Employing Eqs. \eqref{relation} and \eqref{coupled} yields then the effective entry and exit rates,
\begin{equation}
    \alpha_{eff}=\frac{4\omega_a-(2\alpha-1)^2}{4(\omega_a+\omega_d)} , \quad \beta_{eff}=\frac{4\omega_d+(2\alpha-1)^2}{4(\omega_a+\omega_d)}.
    \end{equation}
From these expressions we finally obtain the conditions on the association/dissociation rates as
    \begin{equation}
    \omega_a>\omega_d.
    \end{equation}
 This suggests that this phase can be realized when the association is stronger than the dissociation so that the dynamics in the right sub-lattice becomes limited only by the particle bulk transitions.\\

{\it LD:HD Phase}. In this phase, the left segment is in the LD phase, while the right segment is in the HD phase. For this to happen, the following conditions must be satisfied,
  \begin{equation}\label{ldhdcon}
\alpha\leq\min\{\beta_{eff},0.5\},\qquad \beta\leq\min\{\alpha_{eff},0.5\}
\end{equation}
 The sub-lattice bulk densities can be written as,
 \begin{equation}
\rho_{buk,L_1}=\alpha,\qquad \rho_{bulk,L_2}=1-\beta,
\end{equation}
and the densities at the special sites are
\begin{equation}
\rho_{k-1}=\frac{\alpha(1-\alpha)}{\beta_{eff}},\qquad \rho_{k+1}=1-\frac{\beta(1-\beta)}{\alpha_{eff}}.
\end{equation}
 Plugging these densities into Eq. \eqref{coupled} yields the explicit values for the effective entry and exit rates,
         \begin{eqnarray}
    \alpha_{eff}&=&\frac{\alpha(1-\alpha)+\beta(1-\beta)+\omega_a}{\omega_a+\omega_d} ,\\
    \beta_{eff}&=&\frac{\omega_d-\alpha(1-\alpha)+\beta(1-\beta)}{\omega_a+\omega_d}.
    \end{eqnarray}
It can be further shown that the association and dissociation rates must satisfy the condition
    \begin{equation}
    \omega_a,\omega_d>0.
    \end{equation}
It means that this phase might exist when both association and dissociation rates are not zero.\\

{\it HD:MC Phase} In this phase, the left segment displays the high density, whereas the right segment shows the maximal current. This can happen for
\begin{equation}\label{hdmccon}
\beta_{eff}\leq\min\{\alpha,0.5\},\qquad 0.5\leq\min\{\alpha_{eff},\beta\}.
\end{equation}
The bulk sub-lattice densities  are given by,
 \begin{equation}
\rho_{buk,L_1}=1-\beta,\qquad \rho_{bulk,L_2}=0.5,
\end{equation}
while at the special sites we have,
\begin{equation}
\rho_{k-1}=1-\beta_{eff},\qquad \rho_{k+1}=1-\frac{1}{4\alpha_{eff}}.
\end{equation}
Using the densities $\rho_{k-1}$ and $\rho_{k+1}$ in Eqs. \eqref{coupled} and \eqref{relation}, we compute the effective rates ,
    \begin{equation}
    \alpha_{eff}=\frac{\omega_a}{\omega_a+\omega_d} , \quad \beta_{eff}=\frac{\omega_d}{\omega_a+\omega_d}.
    \end{equation}
Utilising these expressions in the conditions for the existence of this phase given in Eqn. \eqref{hdmccon}, one can easily predict the range of parameters for this phase. In addition, it can be shown that
    \begin{equation}
    \omega_a>\omega_d.
    \end{equation}
Thus, this phase exists when dissociation rates are small enough to support the HD phase in the left sub-lattice, while the association rates are large to ensure the MC phase in the right sub-lattice.\\

\begin{figure*}
 \begin{minipage}{0.45\textwidth}
\vspace*{0.5cm}
%\centering
\includegraphics[width=\textwidth]{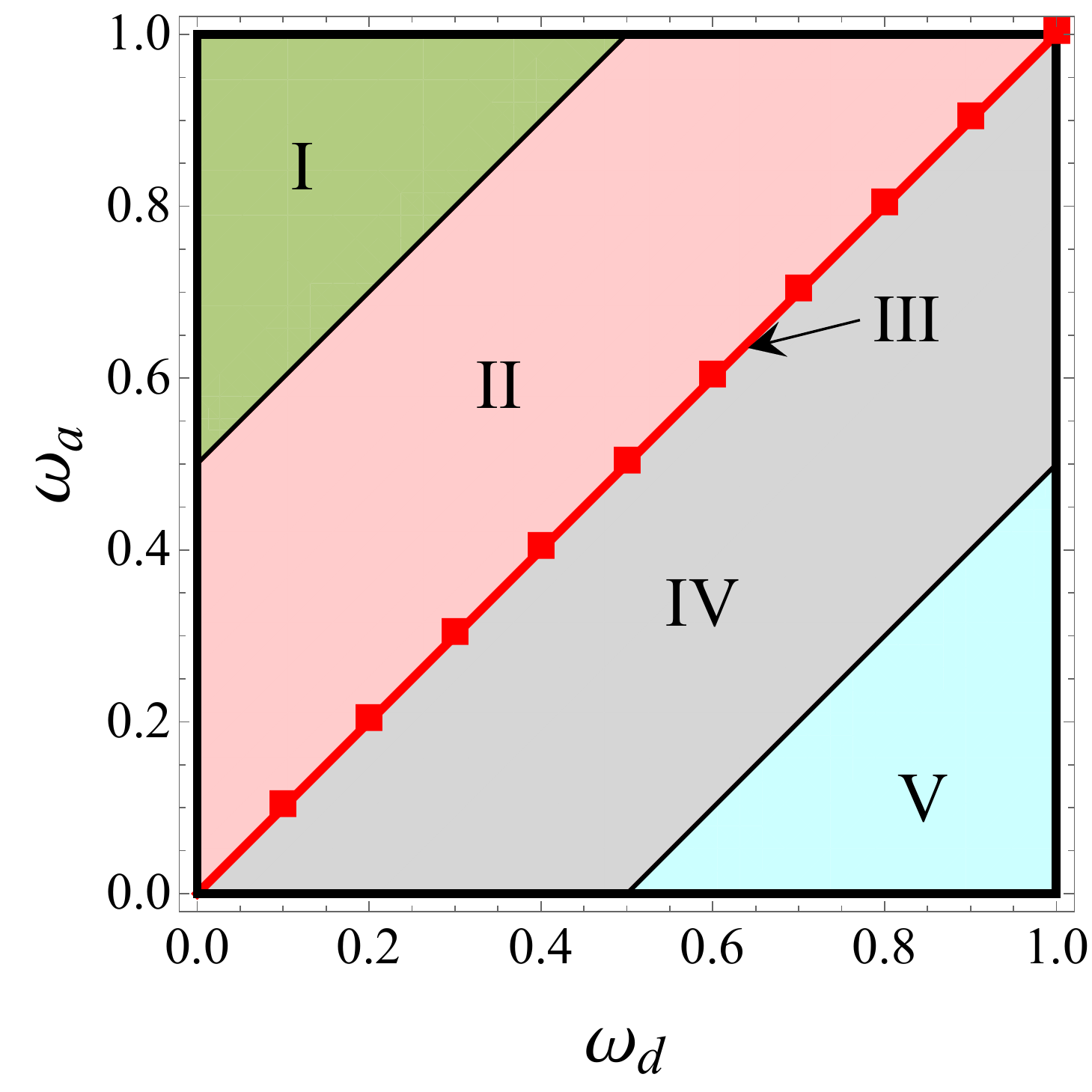}
    \end{minipage}
    %\hfill
   \begin{minipage}{0.45\textwidth}
   \vspace*{0cm}
   % \centering
    \resizebox{0.7\textwidth}{!}{
\begin{tabular}{|c||c|c|c|c|c|}
  \hline
  Phase & I & II & III & IV & V\\
  \hline\hline
  LD:LD & $\times$ & $\checkmark$ & $\checkmark $ & $\checkmark $& $\checkmark$ \\
  LD:MC & $\checkmark$ & $\checkmark$ & $\times$ & $\times$ & $\times$ \\
  LD:HD & $\checkmark$ & $\checkmark$ & $\checkmark$ & $\checkmark$ & $\checkmark$\\
  MC:LD & $\times$ & $\times$ & $\times$ & $\checkmark$ & $\checkmark$\\
  MC:HD & $\times$ & $\times$ & $\times$ & $\checkmark$ & $\checkmark$\\
  MC:MC & $\times$ & $\times$ & $\checkmark$ & $\times$ & $\times$\\
  HD:HD & $\checkmark$ & $\checkmark$ & $\checkmark$ & $\checkmark$ & $\times$\\
  HD:MC & $\checkmark$ & $\checkmark$ & $\times$ & $\times$ & $\times$ \\
  \hline
\end{tabular}
}
\end{minipage}
\caption{\label{topology} Different dynamic regions as a function of the association and dissociation rates. There are five distinct regions labeled as I-V for which the phase regimes differ qualitatively. The phase regimes that exist in these five different possible regions are described in tabular form where ``$\times$" denotes the phase that does not exist..}
\end{figure*}

{\it MC:HD Phase}. This phase has the MC phase in the left sub-lattice and the  HD phase in the right sub-lattice. The conditions for this dynamic regime are the following,
    \begin{equation}\label{mchdcon}
0.5\leq\min\{\alpha,\beta_{eff}\},\qquad \beta\leq\min\{\alpha_{eff},0.5\}.
\end{equation}
The bulk densities in the two segments are,
 \begin{equation}
\rho_{buk,L_1}=0.5,\qquad \rho_{bulk,L_2}=1-\beta,
\end{equation}
and that at the sites $k-1$ and $k+1$ we have,
\begin{equation}
\rho_{k-1}=\frac{1}{4\beta_{eff}},\qquad \rho_{k+1}=1-\frac{\beta(1-\beta)}{\alpha_{eff}}.
\end{equation}
 The calculation of the effective rates produces
    \begin{equation}
    \alpha_{eff}=\frac{4\omega_a+(2\beta-1)^2}{4(\omega_a+\omega_d)}, \quad \beta_{eff}=\frac{4\omega_d-(2\beta-1)^2}{4(\omega_a+\omega_d)}.
    \end{equation}
    Together with the conditions of existence presented in Eqn. \eqref{mchdcon}, one can easily obtain the explicit region of MC:HD phase.
The association/dissociation rates in this phase must satisfy,
    \begin{equation}
    \omega_d>\omega_a.
    \end{equation}
This physically means that this phase might be realized when the dissociation rate is faster than the association rate so that the exit from the system determines the state of the right sub-lattice.\\

\begin{figure*}
\centering
\subfigure[\label{PD0.7}$\Omega=0.7$]{\includegraphics[width=0.3\textwidth]{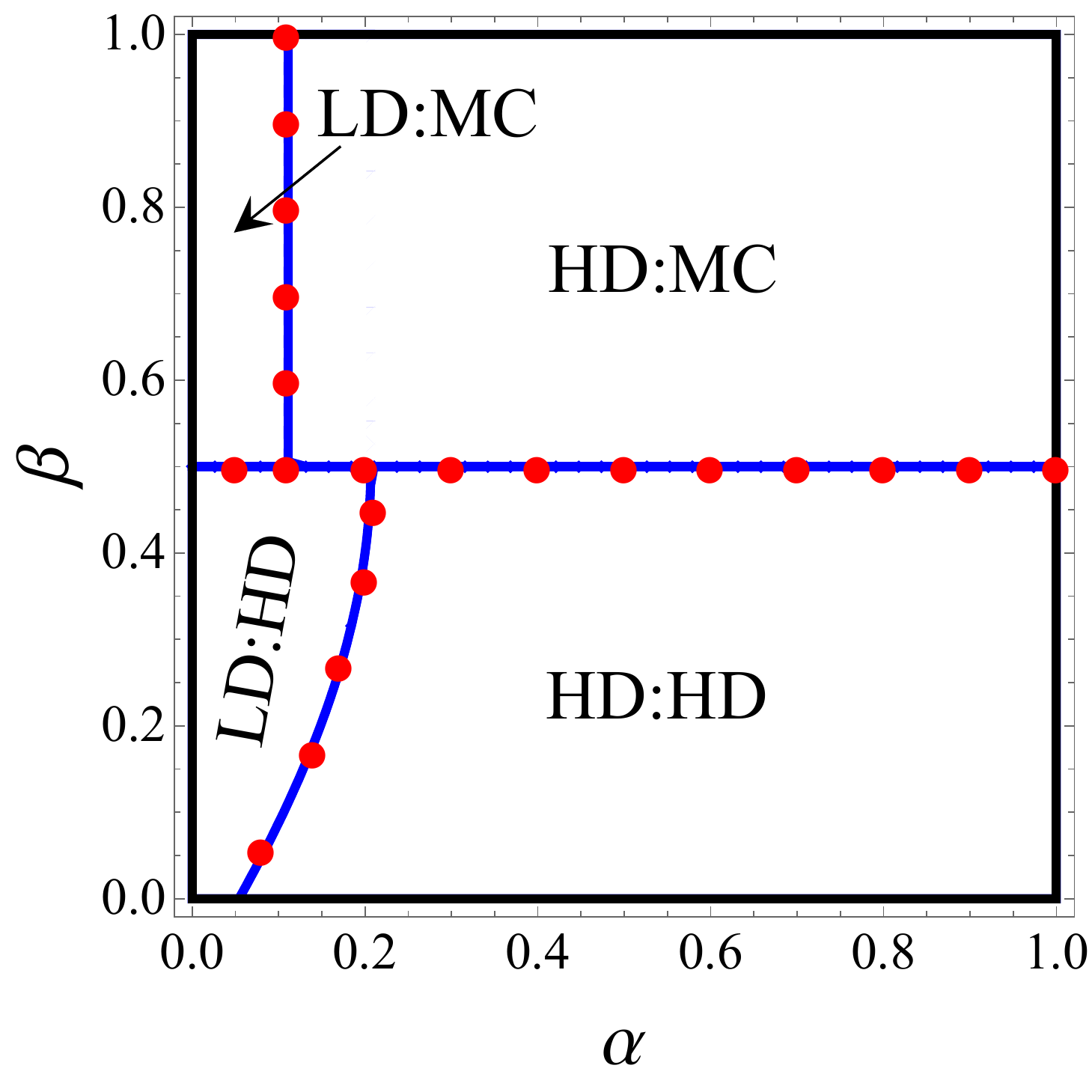}}
\subfigure[\label{PD0.2}$\Omega=0.2$]{\includegraphics[width=0.3\textwidth]{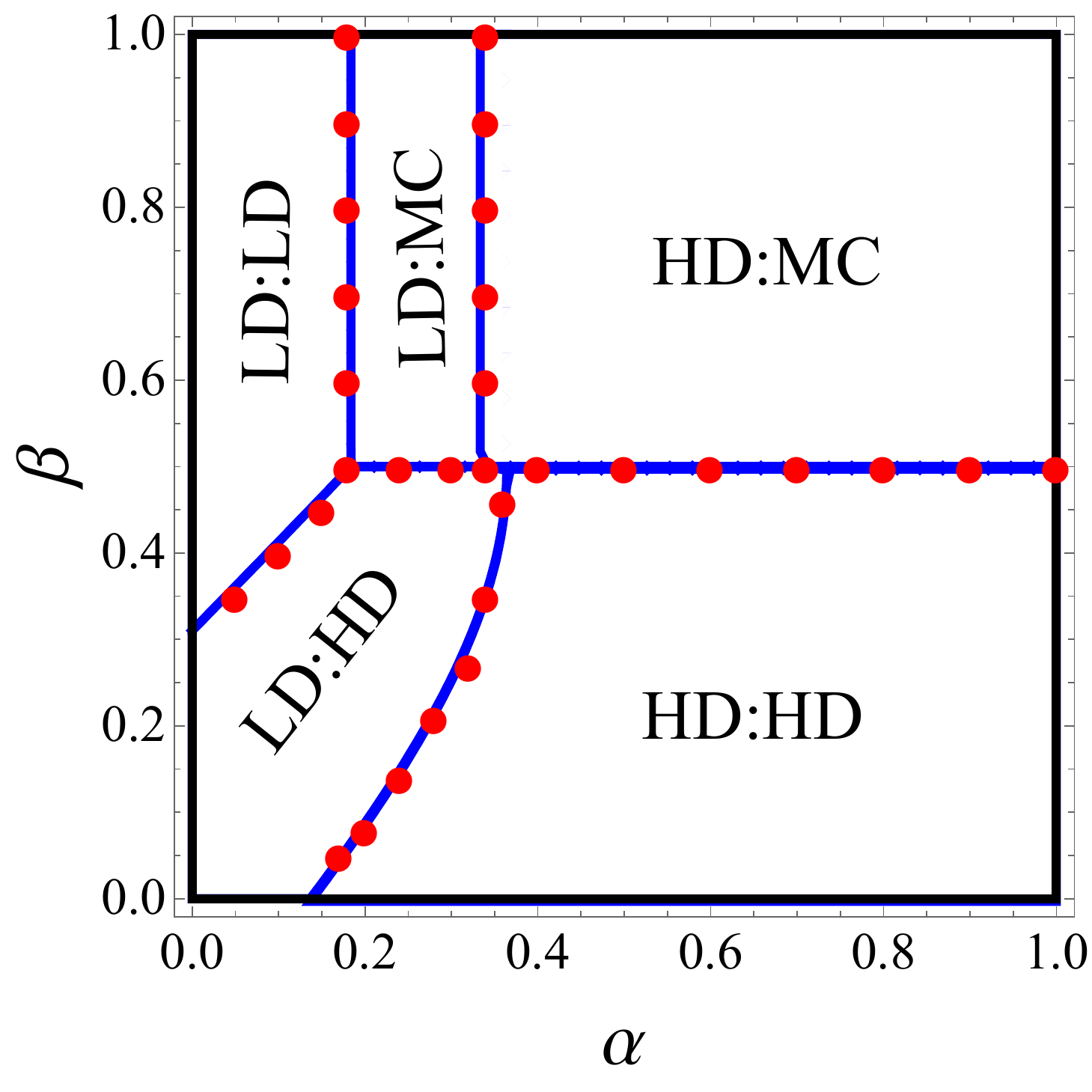}}
\subfigure[\label{PD0}$\Omega=0$]{\includegraphics[width=0.3\textwidth]{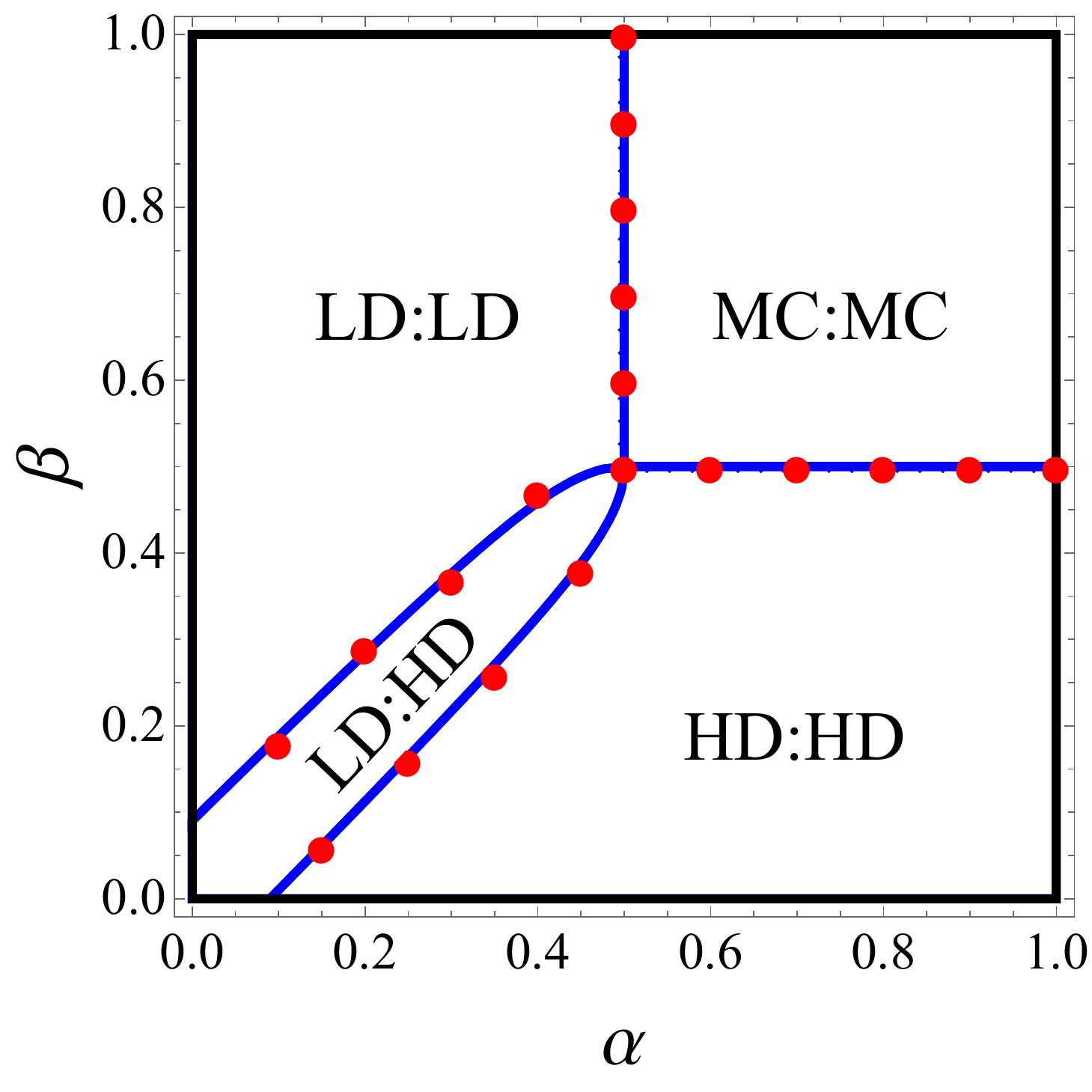}}
\subfigure[\label{PDn0.7}$\Omega=-0.2$]{\includegraphics[width=0.3\textwidth]{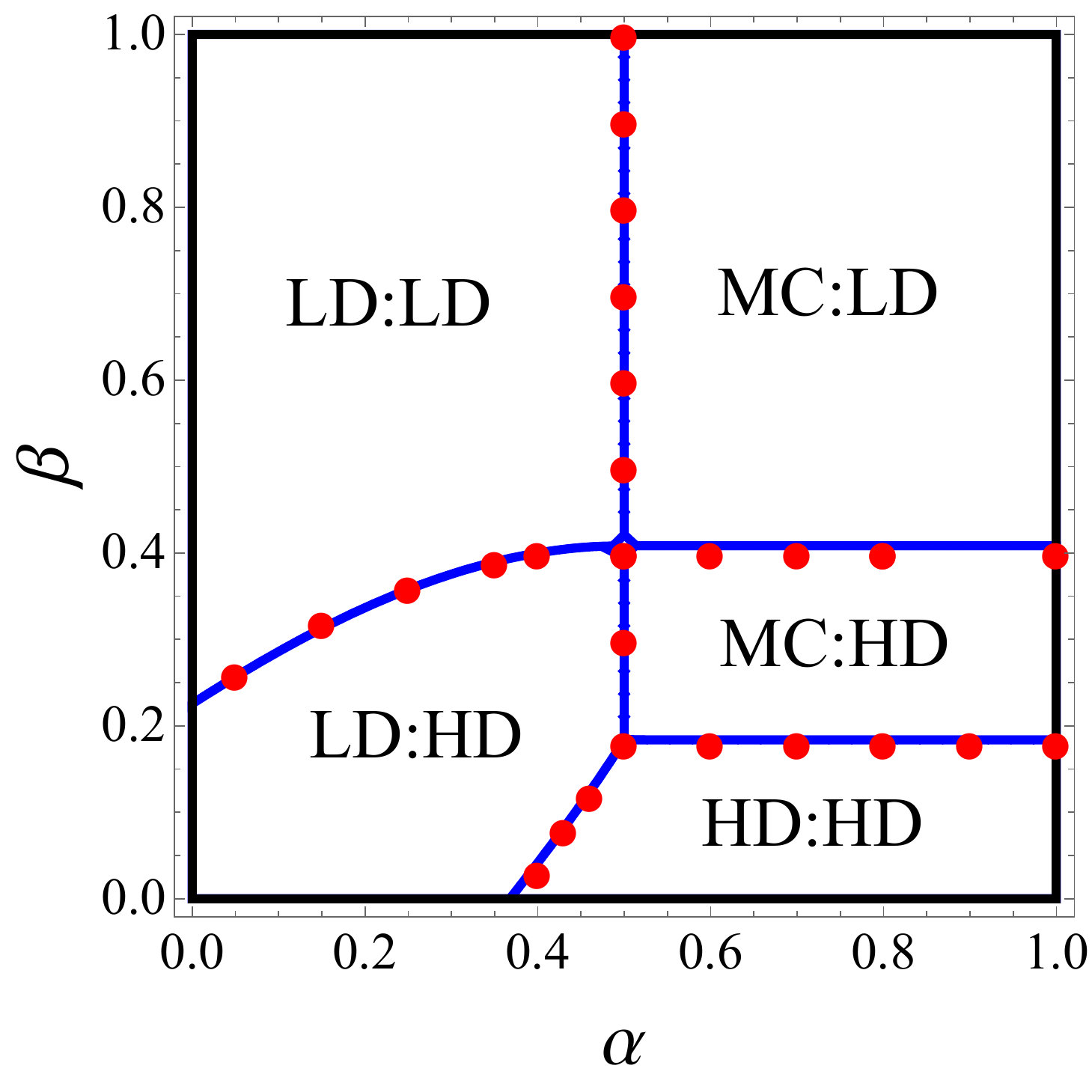}}
\subfigure[\label{PDn0.2}$\Omega=-0.7$]{\includegraphics[width=0.3\textwidth]{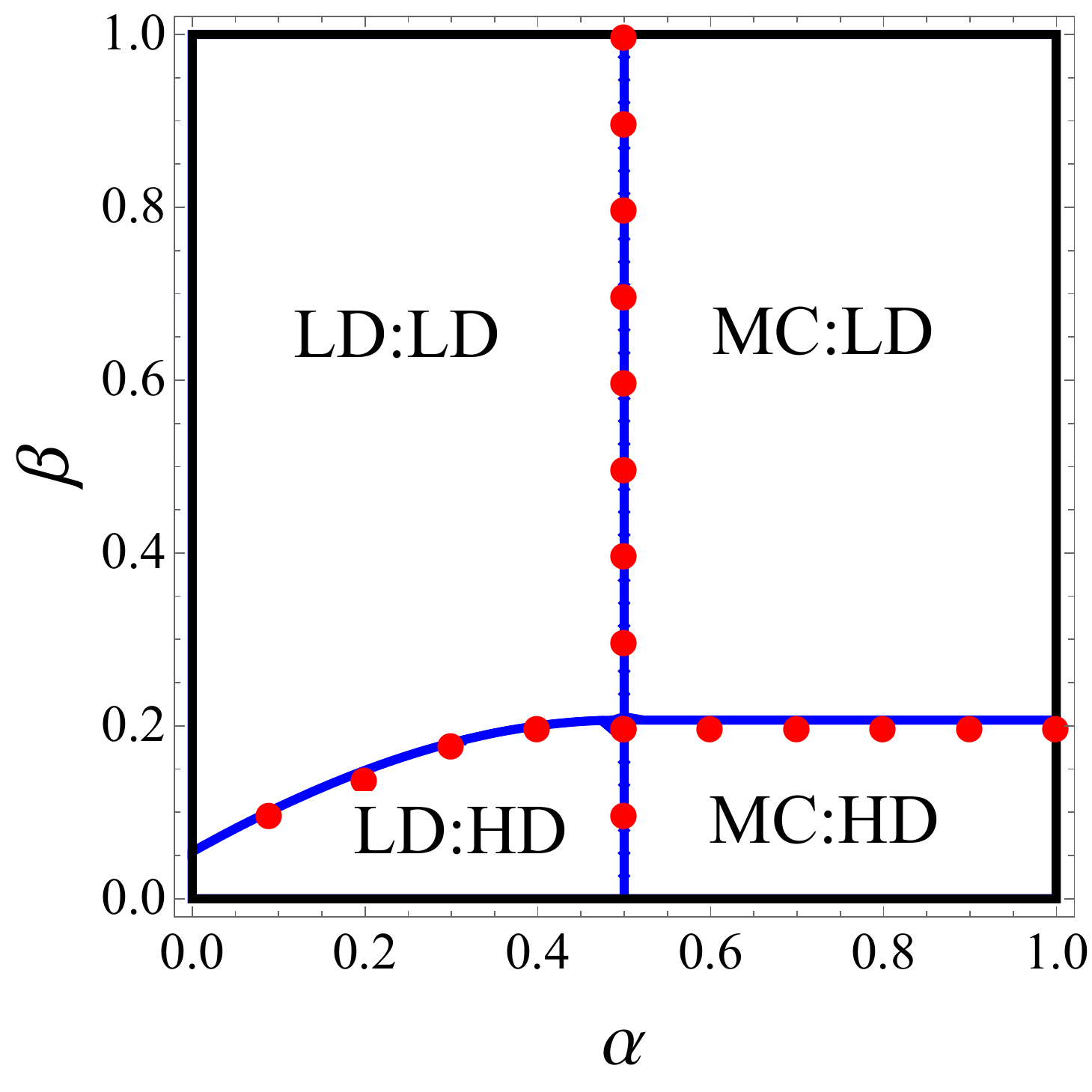}}
\caption{\label{PD}Phase Diagrams in $(\alpha,\beta)$ plane for different values of $\Omega$ notified in sub-captions of figures. (a) $\omega_a=0.8,~\omega_d=0.1$ (b) $\omega_a=0.4,~\omega_d=0.2$ (c) $\omega_a=0.1,~\omega_d=0.1$ (d) $\omega_a=0.4,~\omega_d=0.6$ (e) $\omega_a=0.1,~\omega_d=0.8$.}
\end{figure*}

{\it MC:LD Phase}. In this phase, the left segment  displays the MC phase and the right segment  shows the LD phase. The conditions for this dynamic regime are the following,
     \begin{equation}\label{mcldcon}
0.5\leq\min\{\alpha,\beta_{eff}\},\qquad \alpha_{eff}\leq\min\{\beta,0.5\}.
\end{equation}
The bulk sub-lattice densities are given by,
 \begin{equation}
\rho_{buk,L_1}=0.5,\qquad \rho_{bulk,L_2}=\alpha_{eff},
\end{equation}
while  at the boundaries we have,
\begin{equation}
\rho_{k-1}=\frac{1}{4\beta_{eff}},\qquad \rho_{k+1}=\frac{\alpha_{eff}(1-\alpha_{eff})}{\beta}.
\end{equation}
 Utilising the expression for $\rho_{k-1}$ and $\rho_{k+1}$ in Eq. \eqref{coupled} yields the effective entry and exit rates,
    \begin{eqnarray}
    \alpha_{eff}&=&\frac{1}{2}\left(1+\omega_a+\omega_d-\sqrt{(\omega_a+\omega_d)^2+2(\omega_d-\omega_a)}\right),\hspace{0.6cm}\\
    \beta_{eff}&=&\frac{1}{2}\left(1-\omega_a-\omega_d+\sqrt{(\omega_a+\omega_d)^2+2(\omega_d-\omega_a)}\right) .\hspace{0.6cm}
    \end{eqnarray}
The association and dissociation rates for which this stationary state satisfy
    \begin{equation}
    \omega_d>\omega_a.
    \end{equation}
This phase exists if the dissociation rates are faster than the association rates so that this limits the particle flux into the right sub-lattice,  producing the LD phase in this segment.\\

{\it HD:LD Phase}. In this phase the left and right segments display LD and HD phases, respectively. The conditions that support the existence of this phase are,
     \begin{equation}\label{hdldcon}
\beta_{eff}\leq\min\{\alpha,0.5\},\qquad \alpha_{eff}\leq\min\{\beta,0.5\}.
\end{equation}
The bulk densities in the two segments are
 \begin{equation}
\rho_{buk,L_1}=1-\beta_{eff},\qquad \rho_{bulk,L_2}=\alpha_{eff},
\end{equation}
and the densities at the sites $k-1$ and $k+1$ are given by
\begin{equation}
\rho_{k-1}=1-\beta_{eff},\qquad \rho_{k+1}=\frac{\alpha_{eff}(1-\alpha_{eff})}{\beta}.
\end{equation}
 Now, the conditions that support the existence of this phase from Eq. \eqref{hdldcon} fail to satisfy the Eq. \eqref{relation}. Thus, we predict that this phase cannot exist at any condition. Physically this can be explain using the following arguments.  To keep the left sub-lattice in the HD phase we should have a large association rate. But to keep the LD phase in the right sub-lattice requires very low $\omega_{a}$. The same contradictions exist for the dissociation rates.\\

The conditions for the existence of different stationary phases in terms of the association/dissociation rates are summarized in Table 2. One can see that for non-zero association/dissociation rates there are eight possible stationary dynamic regimes, while in the limiting cases ($\omega_{a}=0$ or $\omega_{d}=0$) the number of possible phases decreases to five. This clearly shows that in order to understand the role of association/dissociation processes in the dynamics of biological molecular motors it is important to consider non-zero association/dissociation fluxes. The limiting situations are not representing the whole complexity of the underlying processes. For example, the MC:MC phase can only be realized when both association and dissociation rates are the same (and non-zero). We should also note that our results in the limiting cases ($\omega_{a}=0$ or $\omega_{d}=0$) fully agree the analysis reported earlier \cite{mirin2003effect,xiao2012single}.

\section{Results and Discussions}

\begin{figure*}
\centering
\subfigure[$\omega_a=0.4,~\omega_d=0.2$]{\includegraphics[width=0.35\textwidth]{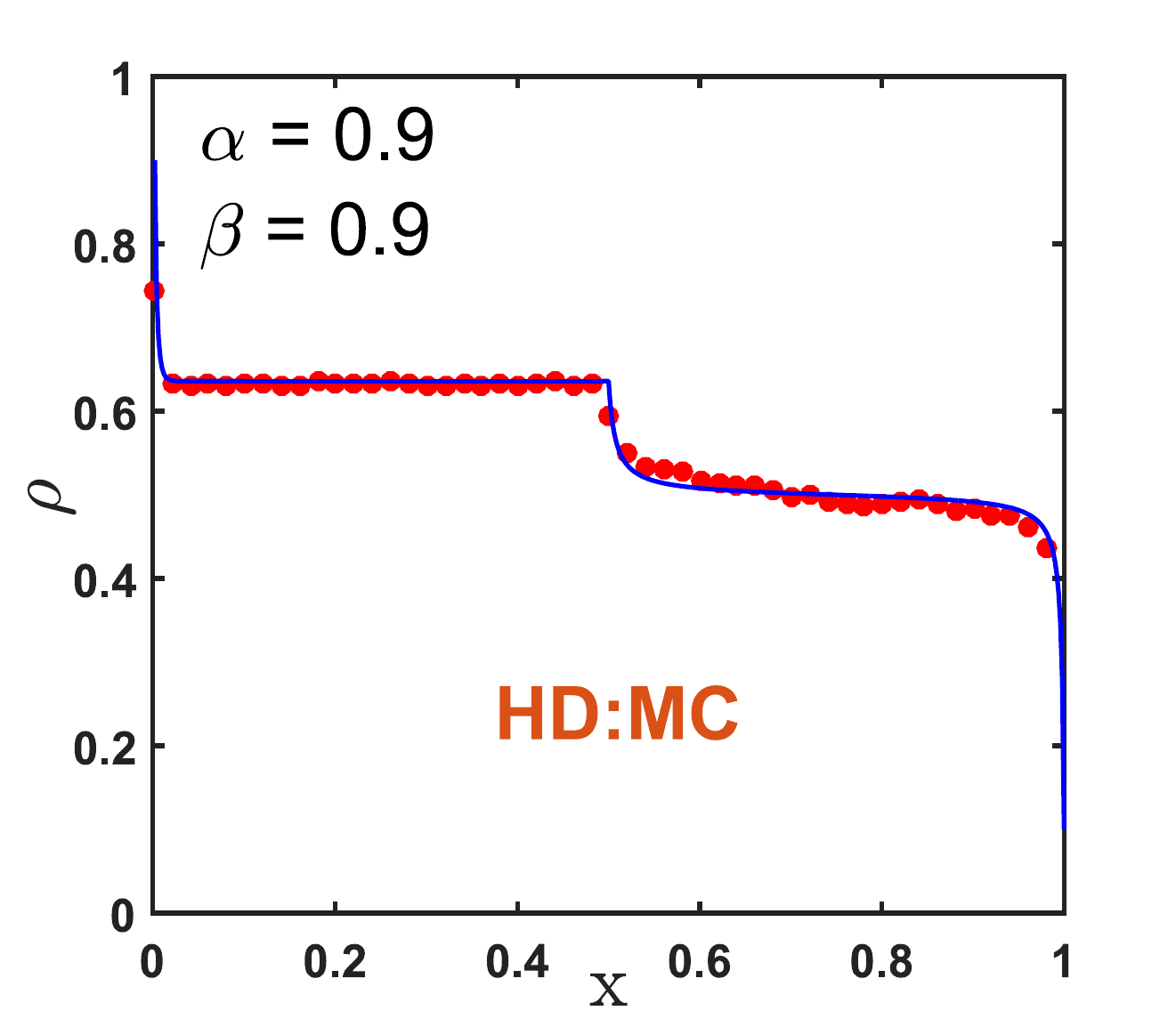}}
\subfigure[$\omega_a=0.4,~\omega_d=0.2$]{\includegraphics[width=0.35\textwidth]{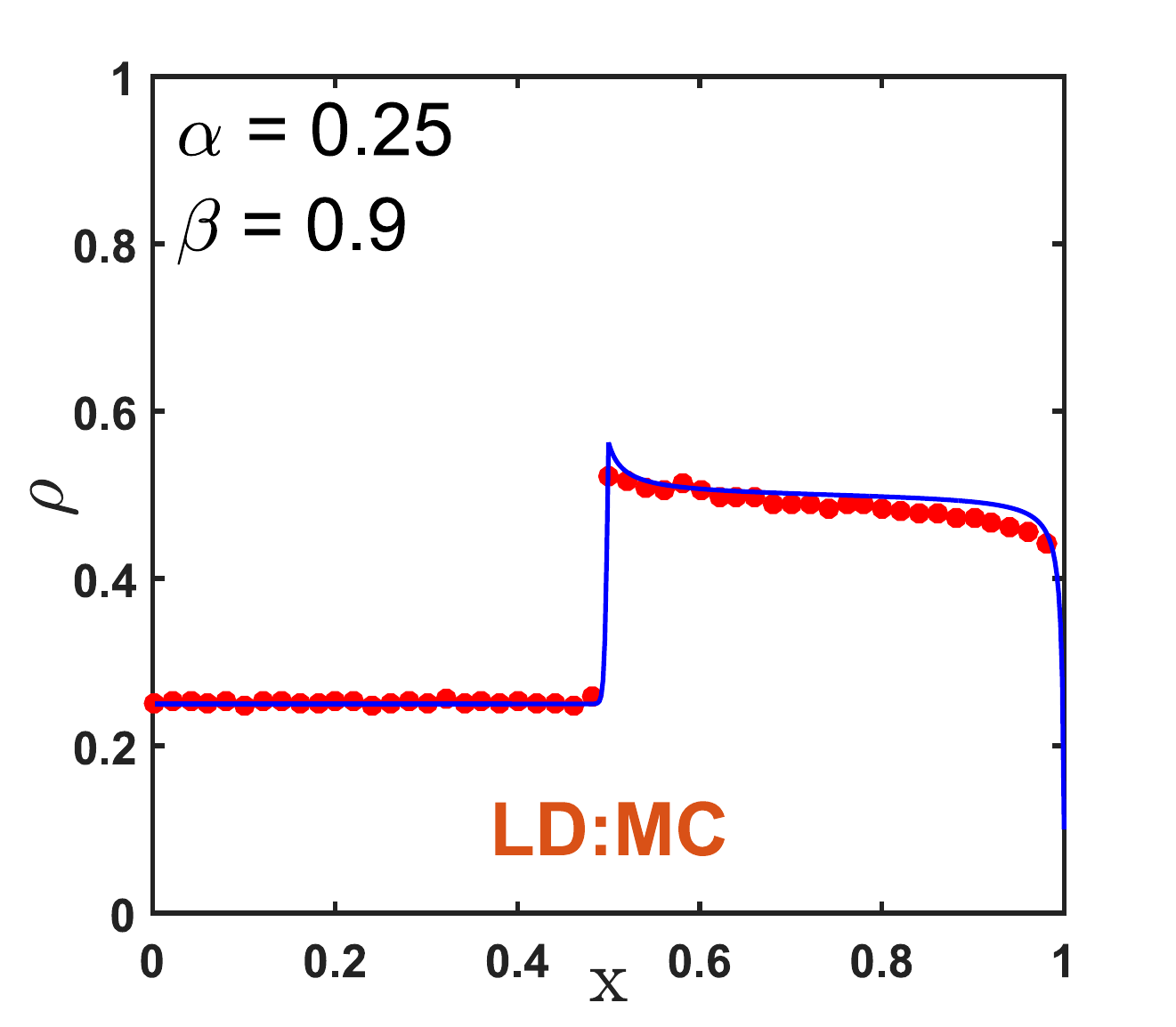}}
\subfigure[$\omega_a=0.4,~\omega_d=0.2$]{\includegraphics[width=0.35\textwidth]{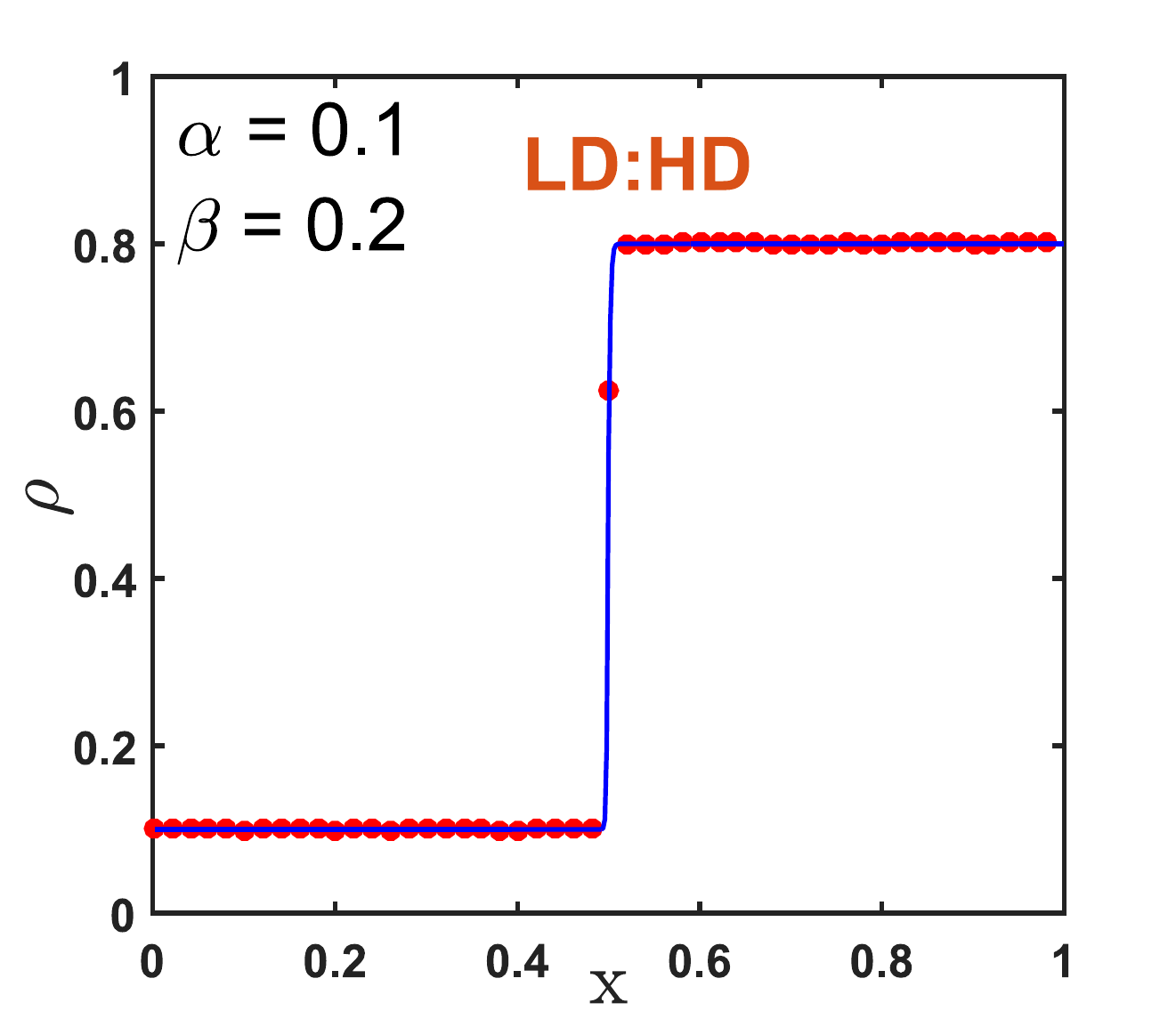}}
\subfigure[$\omega_a=0.4,~\omega_d=0.2$]{\includegraphics[width=0.35\textwidth]{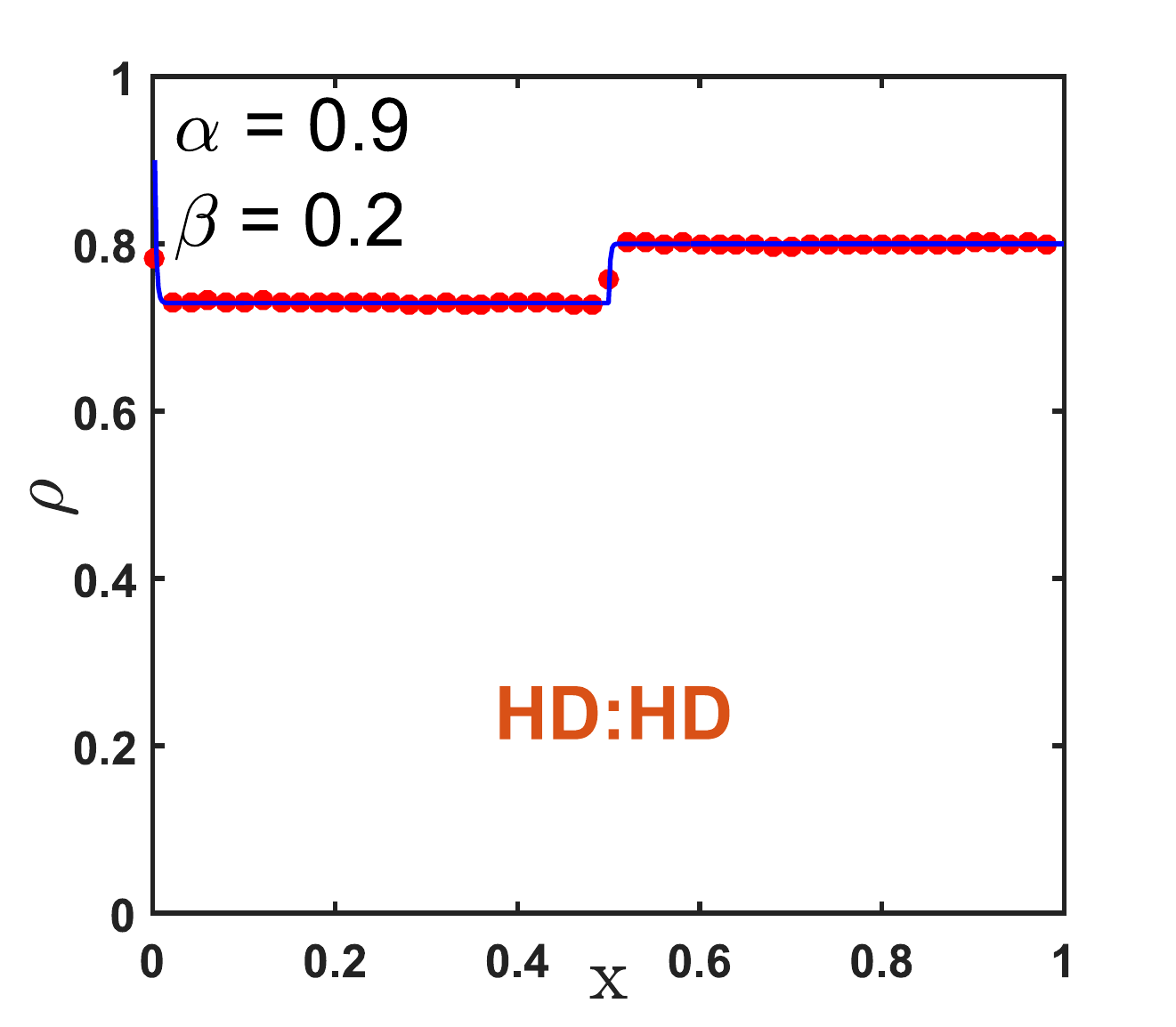}}
\subfigure[$\omega_a=0.4,~\omega_d=0.2$]{\includegraphics[width=0.35\textwidth]{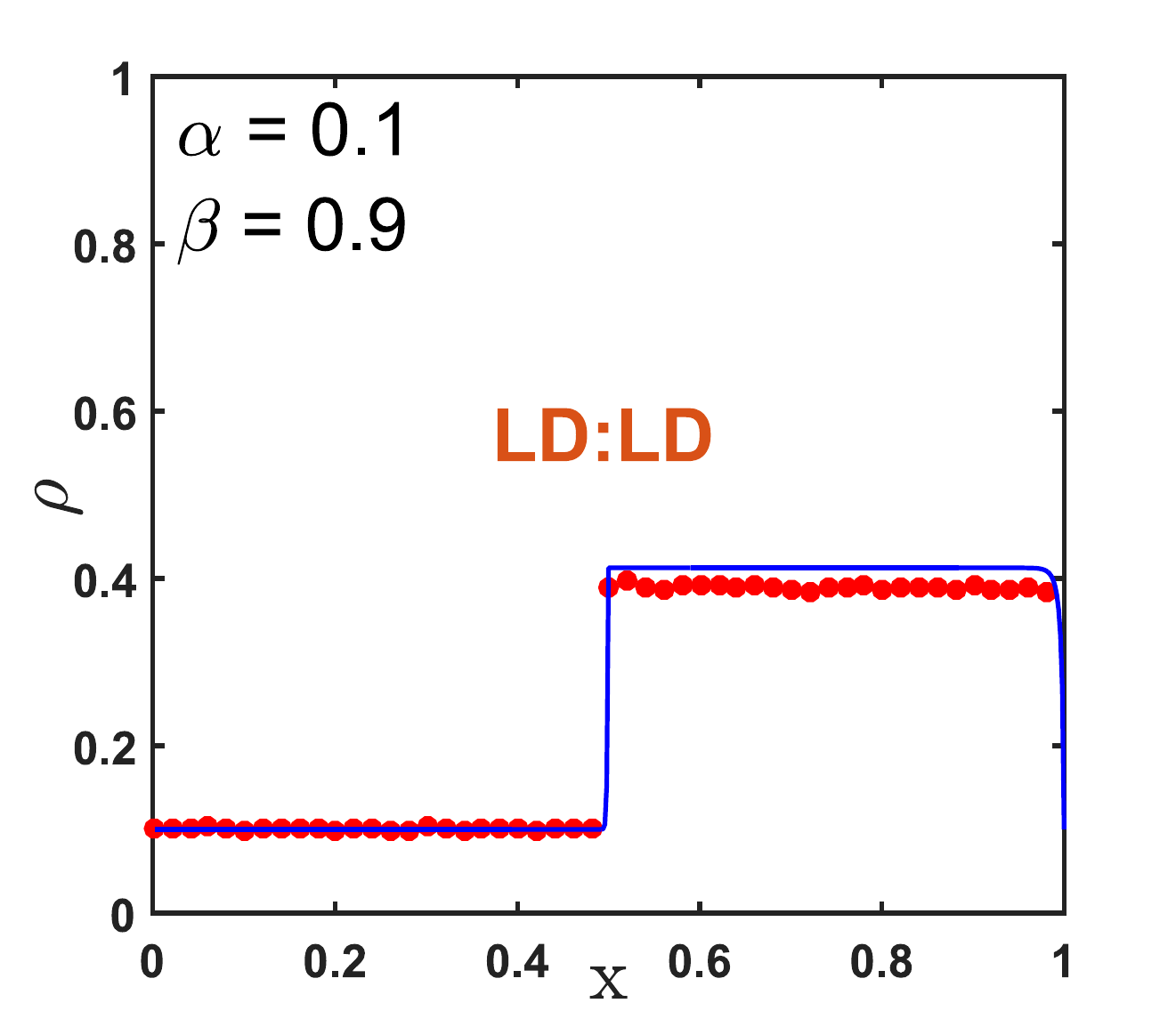}}
\subfigure[$\omega_a=0.4,~\omega_d=0.6$]{\includegraphics[width=0.35\textwidth]{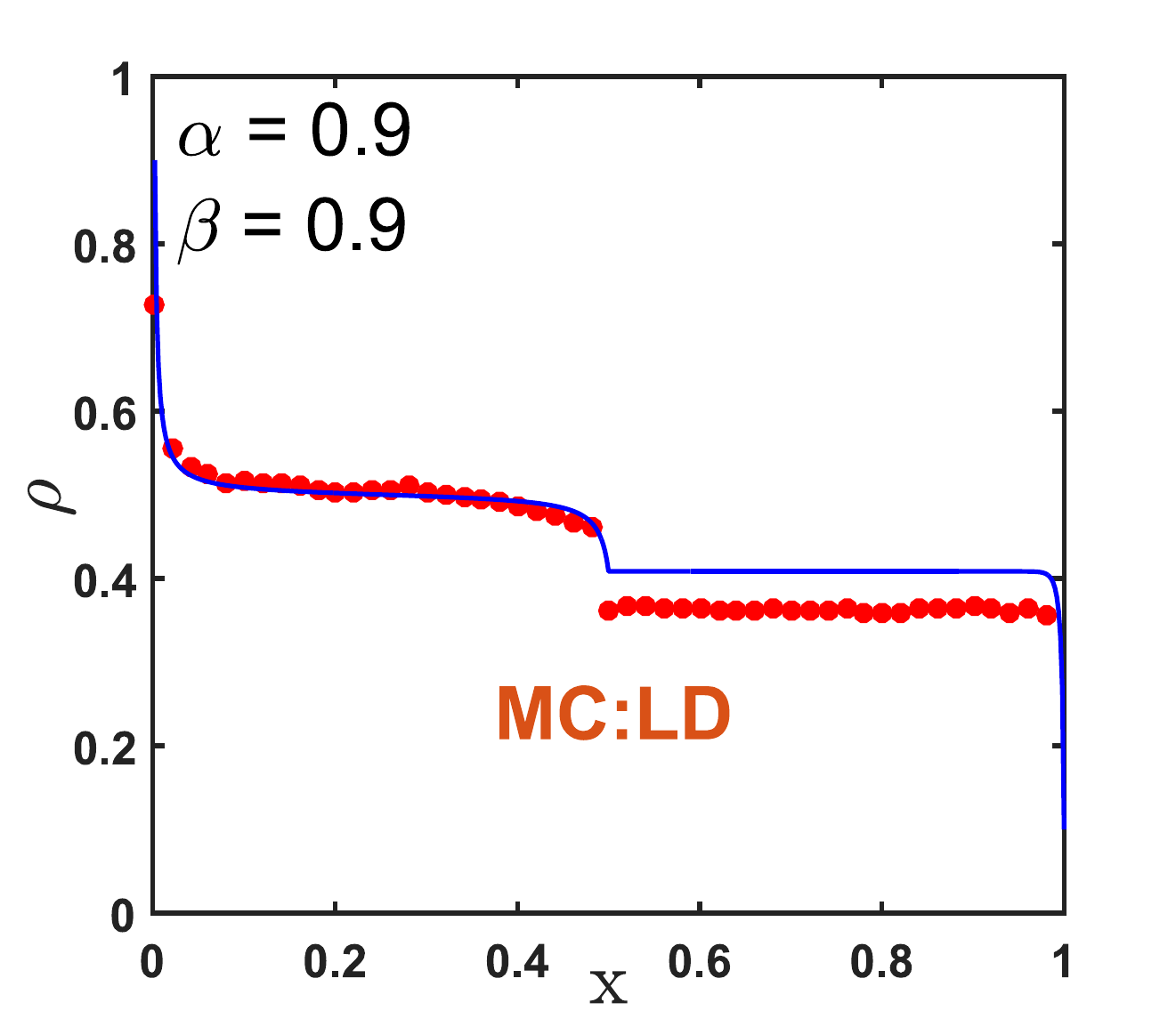}}
\subfigure[$\omega_a=0.4,~\omega_d=0.6$]{\includegraphics[width=0.35\textwidth]{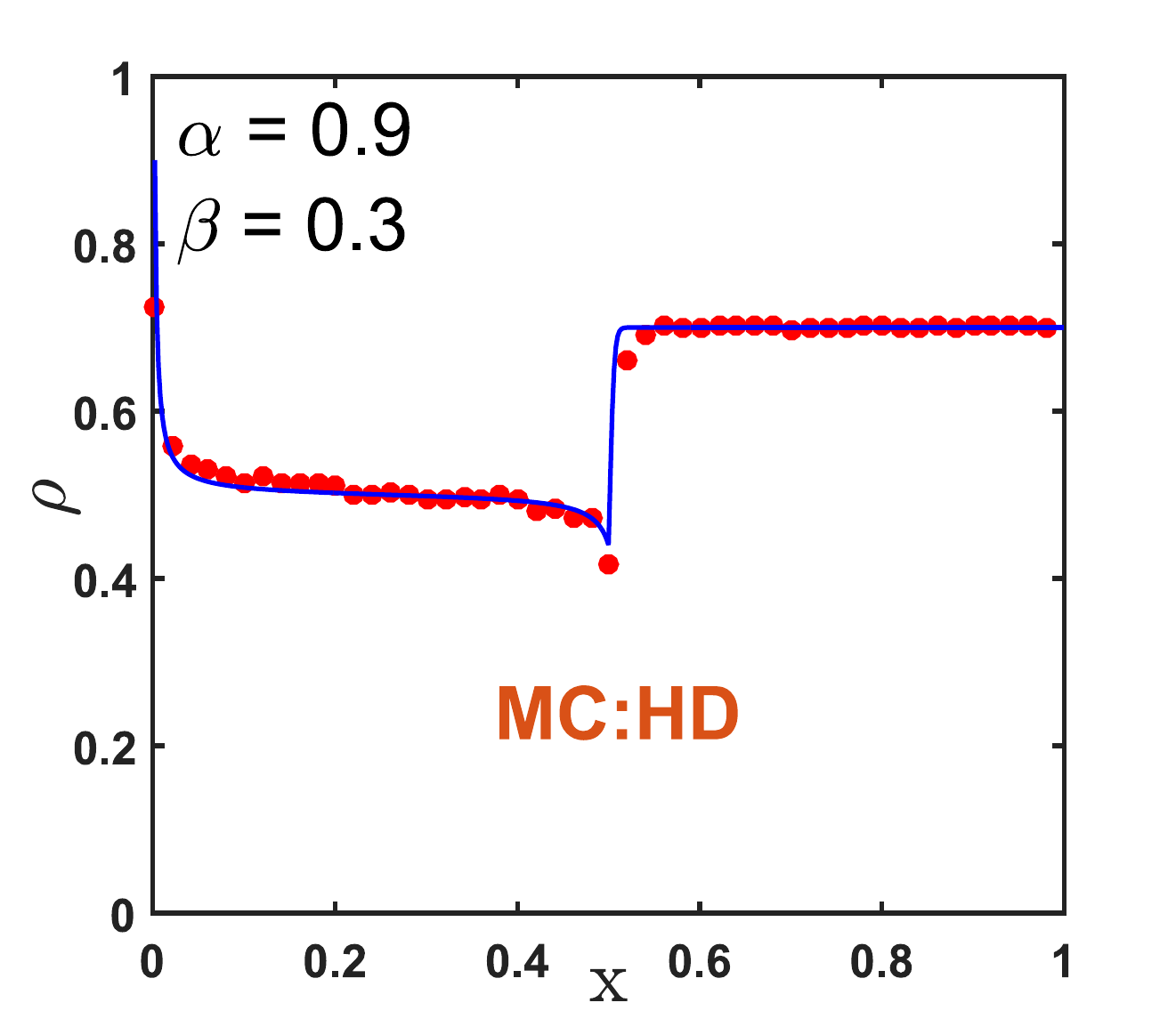}}
\subfigure[$\omega_a=0.1,~\omega_d=0.1$]{\includegraphics[width=0.35\textwidth]{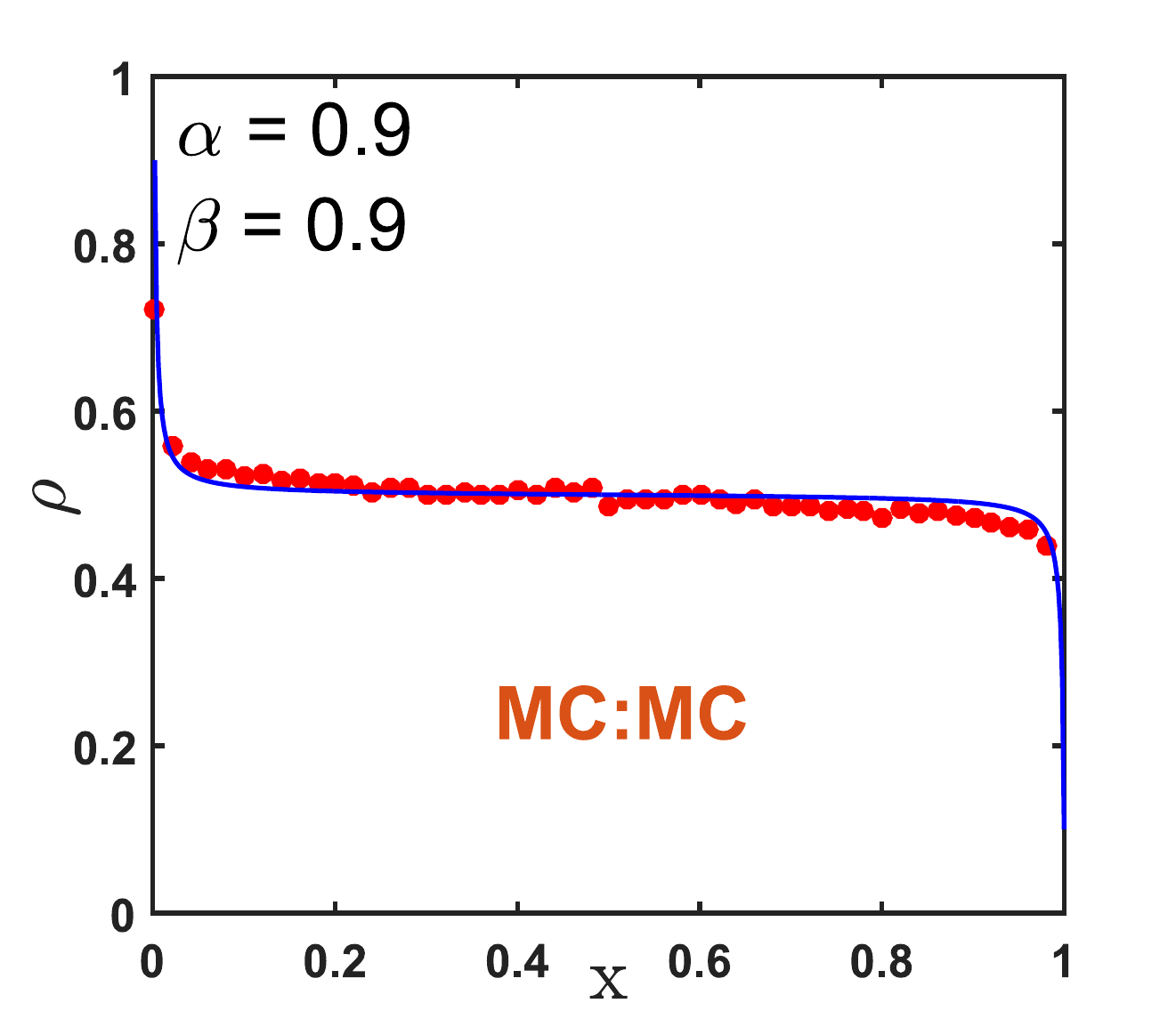}}
\caption{\label{Density}Density profiles in different stationary phases for $\omega_a$ and $\omega_d$ notified in sub-captions and boundary controlling parameters $\alpha$, $\beta$ mentioned in each figure.}
\end{figure*}
Our theoretical approach allows us to fully describe the stationary dynamics of the TASEP model with localized association/dissociation events. But it relies on the sub-lattice mean-field approximation that assumes that the particle densities at the sites $k-1$, $k$ and $k+1$ are independent of each other. To test our theoretical results, we performed extensive Monte Carlo simulations for various ranges of parameters. The simulations are carried out for the system size $N=1000$ following random sequential update rule and are allowed to run for $2\times 10^9$ time-steps to achieve the steady-state conditions. To ensure the occurrence of steady-state, $5\%$ of the initial time-steps are discarded.

Although we predict that up to eight dynamic regimes might be realized in the system, for the fixed association/dissociation rates our analysis of the existence conditions shows that only four or five phases can be observed by varying the global entry ($\alpha$) and exit ($\beta$) rates. This is illustrated in Fig. \ref{topology} where five distinct regions are identified for different association/dissociation rates. To quantify this effect, we introduce a parameter $\Omega=\omega_a-\omega_d$ that lies within the range $[-1,1]$. This parameter describes the difference between the association and dissociation rates. Then these regions correspond to $0.5 < \Omega \le 1$ (region I), $0 < \Omega \le 0.5$ (region II), $\Omega=0$ (region III), $-0.5 \le \Omega < 0$ (region IV) and $-1 \le \Omega < -0.5$ (region V): see Fig. \ref{topology}. To understand what phases will be realized in the system, let us consider the dynamics in all regions in detail.

For the region I ($0.5 < \Omega \le 1$), the association rate is always significantly larger than  the dissociation rate, and this prevents the formation of the LD phase in the sub-lattice $L_2$ because the effective entrance rate into this segment becomes too large. In addition, it prevents the occurrence of the MC phase in the sub-lattice $L_1$ because the particle cannot easily exit the left sub-lattice due to the pile up at the special site $k$. These arguments suggest that only four phases are possible in the region I: LD:MC, LD:HD, HD:HD and HD:MC.

In the region II ($0 < \Omega \le 0.5$), the association rate is only slightly larger than the dissociation rate, and this allows one more additional phase (LD:LD) to be realized. This is because the overall entrance flux into the right sub-lattice is small due to the LD phase in the left sub-lattice and relatively small overall net flux coming into the system from the association/dissociation events. Thus, five phases can be observed in this range of parameters.

In the region III ($\Omega=0$), the association and dissociation rates are equal to each other, and this allows for the MC:MC phase to appear. In this case, the association flux is fully compensated by the dissociation flux and $\rho_{k}=1/2$ [see Eq. (2)].  For low entrance rates $\alpha$, the LD:LD phase also can exist in this region. Similarly, for low exit rates $\beta$, the HD:HD phase can be realized in the system. In addition, the LD:HD phase can be found here too because both LD and HD phases are described by relatively small fluxes that are not affected much by the association/dissociation processes.

The region IV ($-0.5 \le \Omega < 0$) is similar to the region II. Here the dissociation rate is slightly larger than the association rate, and this leads to the possibility of having 5 stationary phases: LD:LD, LD:HD, MC:LD, MC:HD and HD:HD. As expected, it is not possible to have phases with the MC regime on the right sub-lattice, while the MC phase might happen on the left segment.

The dissociation rates are much larger than the association rates in the region V ($-1 \le \Omega < -0.5$). The possible phases here are the same as in the region IV except the phase HD:HD. It cannot exist because the fast removal of particles at the special site prevents the formation of HD phase in the segment $L_1$.

Our theoretical analysis suggests that there are five possible stationary phases in regions II and IV, while other regions have four stationary phases each, as shown in Fig. \ref{PD}. Clearly, we can observe that the stationary phase diagram shows non-monotonic behavior for varying values of $\omega_a$ and $\omega_d$ in regions I-V. Theoretical predictions are fully supported by Monte Carlo computer simulations. It is interesting to note the LD:HD phase can be realized for all possible combinations of association/dissociation rates, while the MC-MC phase can only be found in the symmetric case of $\omega_{a}=\omega_{b}$.

We also predict that there are eight different stationary phase observed in the system, and the corresponding results are presented in Fig. \ref{Density}. One can see that our approximate theory works remarkably well in most situations as compared with Monte Carlo computer simulations. The only deviations are found in the LD:LD and MC:LD phases. These observations can be explained using the following arguments. Our approach assumes that the occupancy of the special site $k$ is independent of the occupancy of the neighboring sites $k-1$ and $k+1$, while in reality some correlations are expected. This would quantitatively affect the effective entrance rates into the right sub-lattice. as the result, the phases with the LD regime in the segment $L_2$ would be affected most by our approximation.

\section{Summary and Conclusions}

A theoretical method to investigate the role of reversible association events in the one-dimensional dynamics of driven particles that interact only via exclusion and when the association/dissociation is localized to a special site far away from the boundaries is presented. The model is motivated by the transport of biological molecular motors moving along linear filaments and their tendency to occasionally reversibly associate to the tracks. Noting that the site of association/dissociation events inserts the inhomogeneity into the system, we approximate the model as two homogeneous segments coupled by the special site. This allows us to obtain a full explicit description of the stationary dynamics and analyze the effect of the localized association/dissociation processes in the particle transport. It is found that from nine possible stationary phase only eight can be realized for different ranges of parameters. Our calculations also show that there are five distinct dynamic regions for varying association and dissociation rates. At each of these regions, only five or four stationary phases can  exist simultaneously. Microscopic arguments to explain these observations are presented. Analytical calculations are tested by extensive Monte Carlo simulations, and excellent agreement is found in most dynamic regimes. We also argue that it is important to have both association and dissociation rates to be non-zero in order to fully understand the complex dynamics of molecular motors that can reversibly associate from the filament.

One could also speculate on possible biological implications of our theoretical results. We can argue that tuning the local association/dissociation rates for motor proteins could lead to significant global dynamic changes, indicating that this might be an efficient method of regulating cellular processes. It will be interesting to test this idea in experiments.

While it seems that the presented theoretical model captures main dynamic features of biological molecular motors that can sporadically dissociate into the solution and return back to the track, it is crucial to discuss the limitations and future directions. The main weakness of our theoretical approach is the neglect of correlations near the special site where association and dissociation events are taking place. In addition, the sizes of biological molecular motors are typically large, occupying more than one lattice site of the underlying filament. Furthermore, there are multiple experimental evidences suggesting that motors can interact with each other and this might strongly modify the dynamics in the system.  It will be important to investigate these possibilities and extensions with more details.

\section*{Acknowledgements}
A.B.K. acknowledges the support from the Welch Foundation (C-1559), from the NSF (CHE-1953453 and MCB-1941106), and the Center for Theoretical Biological Physics sponsored by the NSF (PHY-2019745).

\noindent A.K.G. acknowledges support from DST-SERB,
Govt. of India (Grant CRG/2019/004669).\\

%\bibliographystyle{plain}
%\bibliography{Ref}

\end{document}